\documentstyle[12pt,twoside]{article}
\begin{document}\hbadness=10000\thispagestyle{empty}
\pagestyle{myheadings}
\markboth{H.-Th. Elze and O. Schipper}
{Time without time: a stochastic clock model}
\title{{\bf Time without time: a stochastic clock model}}
\author{$\ $\\
{\bf Hans-Thomas Elze and Otavio Schipper}\\ $\ $\\
Instituto de F\'{\i}sica, Universidade Federal do Rio de Janeiro \\
\,Caixa Postal 68.528, 21945-970 Rio de Janeiro, RJ, Brazil }
\vskip 0.5cm
\date{May 2002}
\maketitle
\vspace{-8.5cm}
\vspace*{8.0cm}
\begin{abstract}{\noindent
We study a classical reparametrization-invariant system, in which ``time'' is 
not a priori defined. It consists of a nonrelativistic particle moving in five 
dimensions, two of which are compactified to form a torus. There, assuming a 
suitable potential, the internal motion is ergodic or more strongly irregular. 
We consider quasi-local observables which measure the system's ``change'' in a 
coarse-grained way. Based on this, we construct a statistical timelike parameter,  
particularly with the help of maximum entropy method and Fisher-Rao information 
metric. The emergent reparametrization-invariant ``time'' does not run smoothly 
but is simply related to the proper time on the average. For sufficiently low 
energy, the external motion is then described by a unitary quantum mechanical 
evolution in accordance with the Schr\"odinger equation.
}\end{abstract}
\section{Introduction}
Motivated by attempts to quantize gravity, based on the classical theory of General Relativity,  
there has recently been interest in the quantization of ``timeless'' reparametrization-invariant 
systems, see, for example, Refs.\,\cite{Rovelli90,Lawrie96,Peres97,Ohkuwa99} and further 
references therein. 
  
Presently, we begin with the study of a classical system and address the question, whether 
local observables can be found which allow to characterize the evolution in a gauge 
invariant way. In previous work it has always been assumed that the global features of the 
trajectories are accessible to the observer, which makes it possible, in principle, to 
express the evolution of an arbitrarily selected degree of freedom ``relationally'' in terms of others 
\cite{MRT99,Montesinos00}. Thereby the Hamiltonian and possibly additional constraints have been eliminated in favour of Rovelli's ``evolving constants of motion'' \cite{Rovelli90}. 
  
In distinction, we presently attempt to characterize the evolution by invariant 
quasi-local statistical properties of the ergodic internal ``clock'' motion. 
Heuristically, we assume ``time is change'' and try to quantify the former in terms of measurements of the latter.  
Our results indicate that a `deparametrized' time evolution can be constructed based on coarse-graining localized  
observations in a classical reparametrization-invariant system which is ergodic. 

We remark that certain forms of globally incomplete statistical knowledge about a classical system lead to its effective quantization locally \cite{Wetterich01}. This points towards a deterministic origin of quantization 
and certainly raises further interesting questions about the relation to the problem at hand. Indeed we shall find 
that the external motion of the particle is described by a (discrete-time) Schr\"odinger evolution.
 
Let us consider a fivedimensional model of a ``timeless'' nonrelativistic particle 
with the action: 
\begin{equation}\label{S}
S=\int\mbox{d}t\;L
\;\;, \end{equation}
where the Lagrangian is defined by: 
\begin{equation}\label{L}
L\equiv\frac{1}{2\lambda}
\left [(\partial_t\vec q)^2-r^2(\partial_t\phi_1)^2-r^2(\partial_t\phi_2)^2\right ]
-\frac{\lambda}{2}\left [r^2\omega^2(\widetilde\phi_1^{\;2}+\widetilde\phi_2^{\;2})
+2r^2\Omega^2\widetilde\phi_1\widetilde\phi_2-E\right ] 
\;\;. \end{equation}
Here $\lambda$ stands for an arbitrary ``lapse'' function of the parameter $t$, 
$\vec q\in\mathbf{R^3}$ denotes an ordinary vector, and $\phi_1$ and $\phi_2$  
denote the angular variables corresponding to the toroidally compactified dimensions with radius 
$r$, respectively; $\omega^2,\Omega^2$ are angular velocity squared coupling parameters. 
The parameter E fixes the energy difference 
between the ``external'' and the compactified ``internal'' degrees of freedom.

Suitably redefining the units of length and energy, we set $r\equiv 1$ henceforth.   
The notation $\widetilde\phi$ indicates that the corresponding terms in $L$ are 
periodically continued: 
\begin{equation}\label{ratchet}
\widetilde\phi\equiv\phi -n\;\;,\;\;\;\phi\in [n,n+1[
\;\;, \end{equation} 
for any integer $n$. Thus, a ratchet type potential results in the $\phi_{1,2}$-plane. 
Alternatively, we may consider the angular variables to be 
normalized to the square $[0,1[\times [0,1[$, of which the opposite boundaries are 
identified, thus describing the surface of a torus with main radii $1/2\pi$.  
    
The relative minus sign between the kinetic terms allows the energy of the 
external motion to be unbounded from above. A similar feature is common to simplified cosmological models 
in accordance with General Relativity, as discussed, for example, in Refs.\,\cite{Peres97,Ohkuwa99} 
and references therein. Furthermore, we choose the potential in the compactified 
dimensions to be unstable, in order to generate a chaotic internal motion which is ergodic and 
possibly more strongly irregular. 
  
Varying the action with respect to $\vec q$, $\phi_1$, and $\phi_2$, we obtain the 
equations of motion: 
\begin{eqnarray}\label{q}
\frac{1}{\lambda}\partial_t(\frac{1}{\lambda}\partial_t\vec q)&=&0
\;\;, \\ [1ex] 
\label{phi1} 
\frac{1}{\lambda}\partial_t(\frac{1}{\lambda}\partial_t\phi_1 )&=& 
(\omega^2\widetilde\phi_1 +\Omega^2\widetilde\phi_2)
(1-\sum_n\delta (\phi_1-n)\;)
\;\;, \\ [1ex] 
\label{phi2} 
\frac{1}{\lambda}\partial_t(\frac{1}{\lambda}\partial_t\phi_2 )&=& 
(\omega^2\widetilde\phi_2 +\Omega^2\widetilde\phi_1)
(1-\sum_n\delta (\phi_2-n)\;)
\;\;, \end{eqnarray}  
where the singular terms arise due to the discontinuities 
of the potential.     
Variation of $S$ with respect to $\lambda$ yields the constraint: 
\begin{equation}\label{constraint} 
\frac{1}{2\lambda^2}
\left [(\partial_t\vec q)^2-(\partial_t\phi_1)^2-(\partial_t\phi_2)^2\right ]
+\frac{1}{2}\left [\omega^2(\widetilde\phi_1^{\;2}+\widetilde\phi_2^{\;2})
+2\Omega^2\widetilde\phi_1\widetilde\phi_2-E\right ] 
=0\;\;, \end{equation}
which will be recognized as a constraint on the Hamiltonian momentarily. 

The canonical momenta are defined as usual: 
\begin{eqnarray}\label{p}
\vec p&\equiv&\frac{\partial L}{\partial (\partial_t\vec q)}=\frac{1}{\lambda}\partial_t\vec q
\;\;, \\ [1ex] 
\label{pi1} 
\pi_1&\equiv&\frac{\partial L}{\partial (\partial_t\phi_1)}=-\frac{1}{\lambda}\partial_t\phi_1
\;\;, \\ [1ex] 
\label{pi2} 
\pi_2&\equiv&\frac{\partial L}{\partial (\partial_t\phi_2)}=-\frac{1}{\lambda}\partial_t\phi_2
\;\;. \end{eqnarray} 
Incorporating these, we obtain the Hamiltonian: 
\begin{eqnarray}\label{H1}
H&=&\vec p\cdot\partial_t\vec q+\pi_1\partial_t\phi_1+\pi_1\partial_t\phi_2-L
\\ [1ex] \label{H2} 
&=&\frac{\lambda}{2}\left [\vec p^{\;2}-\left (\pi_1^{\;2}+\pi_2^{\;2}
-\omega^2(\widetilde\phi_1^{\;2}+\widetilde\phi_2^{\;2})
-2\Omega^2\widetilde\phi_1\widetilde\phi_2\right )-E\right ]
\;\equiv\;\lambda C
\;\;. \end{eqnarray} 
In terms of $C$, Eq.\,(\ref{constraint}) implies the primary constraint: 
\begin{equation}\label{C} 
C=0
\;\;, \end{equation} 
which is a weak equality in the sense 
of Dirac's formalism of constraint systems \cite{Dirac}.   
Consequently, the Hamiltonian presents a weak constraint, $H=0$. 

Finally, in Hamiltonian form, the equations of motion (\ref{q})--(\ref{phi2}) read: 
\begin{eqnarray}\label{Hq}
\partial_t\vec p&=&0 
\;\;, \\ [1ex] 
\label{Hphi1}
\partial_t\pi_1&=&-\lambda 
(\omega^2\widetilde\phi_1 +\Omega^2\widetilde\phi_2)
(1-\sum_n\delta (\phi_1-n)\;)
\;\;, \\ [1ex] 
\label{Hphi2} 
\partial_t\pi_2&=&-\lambda 
(\omega^2\widetilde\phi_2 +\Omega^2\widetilde\phi_1)
(1-\sum_n\delta (\phi_2-n)\;)
\;\;. \end{eqnarray} 
Employing these equations and the definition of the constraint $C$ in Eq.\,(\ref{C}), it 
follows by explicit calculation that it does not evolve. Consistently, this is also 
obtained by:  
\begin{equation}\label{Hevol}  
\partial_tC=\{ C,H\}=\lambda^{-1}\{ H,H\}=0
\;\;, \end{equation}
employing the Poisson bracket notation, $\{ A,B\}\equiv \sum (\partial_QA\partial_PB-\partial_PA\partial_QB)$, where a sum over all coordinates and canonical 
momenta, represented by $Q$ and $P$, is understood.
Therefore, no secondary constraints exist in our model. 

We conclude that the system has four physical degrees of freedom. Its extended phase space, cf. Ref.\,\cite{Rovelli90}, is tendimensional, corresponding to the Lagrangian variables  
in Eq.\,(\ref{L}) and the associated canonical momenta. It is, however, 
reduced to a ninedimensional surface by the constraint. Since the physical phase 
space is eightdimensional, there must be one ``gauge'' degree of freedom, which is 
related to the reparametrization invariance. The study of this gauge symmetry and 
its consequences will be performed in Section\,2\,, while the gauge invariant 
description of the evolution will be developed in Section\,3\,. In Section\,4 we 
demonstrate that under certain conditions the external motion of the particle can be 
mapped onto an evolution according to the Schr\"odinger equation. We conclude with a brief discussion.  
 
\section{Gauge invariance and observables}   
We observe indeed that the action is invariant under the set of gauge transformations: 
\begin{equation}\label{gauge} 
t\equiv f(t')\;\;,\;\;\; x(t)\equiv x'(t')\;\;,\;\;\; \lambda (t)\frac{\mbox{d}t}{\mbox{d}t'}\equiv\lambda'(t')
\;\;, \end{equation}
where $x\in\{\vec q,\phi_1,\phi_2\}$.   
Corresponding infinitesimal transformations are generated by: 
\begin{equation}\label{inft} 
\delta t\equiv t-t'=\epsilon (t')
\;\;, \end{equation}
where $\epsilon$ is infinitesimal. This yields immediately:
\begin{eqnarray}\label{infx}
\delta x&\equiv&x(t')-x'(t')=-\epsilon (t')\partial_{t'}x(t')
\;\;, \\ [1ex]
\label{inflambda}
\delta\lambda&\equiv&\lambda (t')-\lambda '(t')=-\partial_{t'}[\epsilon (t')\lambda (t')]
\;\;. \end{eqnarray} 
Employing the definitions of the canonical 
momenta, Eqs.\,(\ref{p})--(\ref{pi2}), we obtain from Eq.\,(\ref{infx}): 
\begin{eqnarray}\label{infq} 
\delta\vec q&=&-\epsilon\lambda\vec p
\;\;, \\ [1ex] 
\label{infphi12}
\delta\phi_{1,2}  &=&\epsilon\lambda\pi_{1,2}    
\;\;. \end{eqnarray}   
Similarly, with the help of Eqs.\,(\ref{Hq})--(\ref{Hphi2}), we obtain: 
\begin{eqnarray}\label{infp}
\delta\vec p &=&0 
\;\;, \\ [1ex]
\label{infpi1}
\delta\pi_1  &=&\epsilon\lambda  
(\omega^2\widetilde\phi_1 +\Omega^2\widetilde\phi_2)
(1-\sum_n\delta (\phi_1-n)\;)
\;\;, \\ [1ex]
\label{infpi2}
\delta\pi_2  &=&\epsilon\lambda 
(\omega^2\widetilde\phi_2 +\Omega^2\widetilde\phi_1)
(1-\sum_n\delta (\phi_2-n)\;)
\;\;. \end{eqnarray} 
Comparing with the equations of motion, we see that the evolution of the coordinates 
and momenta is generated by the gauge transformations. 

Obviously, the three-momentum $\vec p$ is 
conserved, in accordance with translation invariance. Its components present 
three gauge invariant observables which may serve as coordinates of the physical phase space. 
Another invariant  
is provided by the internal contribution to the constraint:
\begin{equation}\label{Cint}
C_{int}\equiv\vec p^{\;2}-E-2C
\;\;, \end{equation}
cf. Eqs.\,(\ref{H2})--(\ref{C}); since the constraint does not evolve, i.e. is invariant, 
as well as $\vec p^{\;2}$, this also holds for $C_{int}$. 
However, being related via the constraint, $C_{int}$ is not  
independent of $\vec p$. 
The angular momentum suggests itself as a 
further observable,  
\begin{equation}\label{J}
\vec J\equiv\vec q\times\vec p
\;\;, \end{equation} 
due to rotational invariance. 
Clearly, $\delta\vec J=\delta\vec q\times\vec p+\vec q\times\delta\vec p =0$. 

As expected, the external motion 
plays a rather passive role in our system, since it can be almost completely described in terms of the  
conserved linear and angular momenta. It only contributes with its kinetic energy  
to the constraint $C$. However, note that $\vec p$ and $\vec J$ together 
determine only the two constant components of $\vec q$ transverse to $\vec p$, 
i.e. $\vec q_\perp$. 
Therefore, in order to predict the  
evolution of the coordinate $\vec q$, we need to construct a 
reference ``clock'', such that the longitudinal component of Eq.\,(\ref{p})  
is well determined, when integrated:
\begin{equation}\label{qevol}
\vec q_\parallel(t)=\vec q_\parallel^{\;(0)}+\vec p\int_0^t\mbox{d}t'\lambda(t')
\;\;. \end{equation}
This is one of the objectives of the following section.    

Furthermore, we remark that we so far obtained five gauge invariant coordinates 
($\vec p$, $\vec q_\perp$) for the eightdimensional physical phase space plus  
one constraint ($C$, or $C_{int}$), which allows to eliminate one more of the 
remaining four internal variables ($\phi_{1,2}, \pi_{1,2}$). Due to the intrinsic 
nonlinearity of the gauge tranformations, Eqs.\,(\ref{infq})--(\ref{infpi2}), 
we are unable to find further invariants in the general, interacting case. 
This motivates our attempt to construct other, statistical observables, in order to 
obtain an invariant, even if coarse-grained description of the physical phase space.   

\section{Ergodicity and coarse-grained ``timing''} 
In order to proceed, we make the crucial assumption that our model forms an ergodic system \cite{Schuster}. We also restrict the allowed lapse functions $\lambda$ to be (strictly) positive, thus avoiding trajectories which trace themselves backwards (or stall) \cite{Lawrie96}.          

In the following, we will consider quantities $N[\phi_{1,2}](c_i)$  
which are functionals only of the trajectory determined by the coordinates $\phi_{1,2}$ and possibly depend parametrically on further constraints involving only them. While some explicit examples will be studied in detail, generally speaking, we have quantities in mind which reflect properties of Poincar\'e sections of the full trajectories. Naturally, we require the constraints $c_i$ to transform as the coordinates under the gauge transformations (\ref{gauge}), i.e. $c_i'=c_i$. It follows that such quantities are gauge invariant, since they depend only on 
geometrical properties of a path and related constraints. Being independent of the momenta, they do not depend on how the trajectory is parametrized: $N[\phi_{1,2}'](c_i')=N[\phi_{1,2}](c_i)$. Thus, they qualify as coarse-grained 
observables characterizing the internal motion. 

Our aim is to construct a timelike variable based 
on such observables. In the following first subsection we do this based on the 
idea that the geometric path length covered by the system evolving from an initial 
to a final state is an invariant measure of the ``time'' that passed. The crucial point is 
that this measure can be inferred in an ergodic system approximately from coarse-graining  localized observations, 
provided we understand the dynamics sufficiently well.    
  
In the second subsection, however, considering the interacting nonlinear system, 
we generalize 
this approach, in order to extract a ``time'' from quasi-local measures of 
the ``change'' occurring while the system evolves. In particular, we will employ 
a maximum-entropy method together with the Fisher-Rao information metric, in order 
to characterize the distance, i.e. the ``time'' passed, between evolving probability 
distributions. We point out that our approach is somehow orthogonal to the one of Ref.\,\cite{Ariel01},   
although we make use of its formalism. There the author launches the ambitious  
project to derive dynamics from rules of inference and the maximum-entropy principle in 
particular. We instead assume the reparametrization-invariant dynamics to be given and construct a 
pertinent notion of ``time''. 

\subsection{Free internal motion on the torus}
Illustrating our approach, we begin with the noninteracting case, i.e., with  $\omega^2=\Omega^2=0$. 
Even without interactions, the internal motion is ergodic for almost all initial conditions.  
In particular, if the 
ratio of the two independent angular velocities is not a rational number, then, for 
sufficiently late times $t$, the trajectory will come arbitrarily close to any point on 
the surface of the torus. This is easily seen in its parametric representation, 
Eq.\,(\ref{phirelat}) below. In \underline{Fig.}\,1 we show a typical example. 

The equations of motion immediately yield the solutions:  
\begin{equation}\label{phifree} 
\phi_{1,2}(t)=\phi_{1,2}^{(0)}-\pi_{1,2}^{(0)}\int_0^t\mbox{d}t'\;\lambda (t') 
\;\;, \end{equation}
where $\phi_{1,2}^{(0)}$ and $\pi_{1,2}^{(0)}$ denote the initial coordinates and 
momenta, respectively; of course, $\pi_{1,2}(t)=\pi_{1,2}^{(0)}$. Eliminating the 
integral of the lapse function, we obtain, for example, $\phi_2$ in terms of $\phi_1$:  
\begin{equation}\label{phirelat} 
\phi_2(\phi_1)=\phi_2^{(0)}-\frac{\pi_2^{(0)}}{\pi_1^{(0)}}(\phi_1^{(0)}-\phi_1)  
\;\;. \end{equation} 
Following Refs.\,\cite{Rovelli90,MRT99}, this is a gauge invariant ``relational'' description 
of the motion: $\phi_2(s)$ gives the value of coordinate $\phi_2$, when (`the time is such   
that') $\phi_1$ has the value $s$. In this way, $\phi_1$ may serve as a time variable, 
even if not a unique one. 

However, similarly as in the models studied previously, 
there is additional information about the full path, which is necessary to complete this description. In the present case, the solution (\ref{phirelat}) presents a line in the $\phi_{1,2}$-plane, unwrapping the motion which multiply covers the torus. Thus, when folding it back onto the torus, one has to keep track of which unit square in the plane a respective piece is coming from. This can be labelled by two 
integers $n_{1,2}$, which may be interpreted as the winding numbers characterizing 
additionally a given point $(\widetilde\phi_1,\widetilde\phi_2)$ on the path. Obviously, 
this presents highly nonlocal (topological) information, which will be unavailable for a local observer under 
more realistic circumstances, such as in the presence of nonlinear interactions.   
       
Therefore, we turn to statistical measures of the motion on the torus. 
Consider the ``incident number'' $I_1^{\delta}(L)$ which counts the number of times that a given trajectory 
$(\widetilde\phi_1(t),\widetilde\phi_2(t))$, cuts the $\phi_1$-axis 
in the neighbourhood of the observer, i.e., with $\widetilde\phi_1\in [0,\delta <1]$ and $\widetilde\phi_2=0$,    
subject to the constraint that the total path length equals $L$, 
taking into account the wrapping around the torus. 
Due to ergodicity and assumed nonnegative lapse functions,  
$I_1$ is a stepwise increasing function of $t$, while $\widetilde\phi_{1,2}(t)\in [0,1[$ are of  
sawtooth type, with details depending on the particular parametrization. The 
reparametrization-invariant value of  
$I_1$, however, only depends on the path length; thus, it implicitly involves nonlocal 
information.  

The incident number $I_1$ presents a very simple example of the coarse-grained observables discussed 
before. There exists an unlimited number of different such counting observables. 
The more of them we introduce and measure, the more detailed will be our reparametrization-invariant 
description of the internal motion. 
We will also use the 
incident number $I_2$ 
which is defined like $I_1$, however, with the roles of coordinates ``1'' and ``2''  
exchanged. 

Furthermore, we may consider one incident number 
as a parametric function of the other:  
\begin{equation}\label{I21}
I_{21}(k)\equiv\mbox{max}\;I_2\;\;,\;\;\mbox{with}\;\;I_1^{\delta}(L)=k 
\;\;, \end{equation} 
where $\delta$ is fixed, while the total path length $L$ is the implicit common variable. 
Note that $I_{21}$ is unique, since we take the respective maximum of $I_2$, which 
may increase while $I_1$ stays constant temporarily. 
 
However, the constraint on a given path length is irrelevant for the actual values assigned to $I_1$ 
and $I_{21}$. One operationally determines them   
by counting the localized incidents, with $\widetilde\phi_{1,2}(t)\in [0,\delta]$, 
and recording one incident number as a function of the other; no knowledge of the  
path length is required. 

Nevertheless, these reparametrization-invariant numbers can be used to determine  
the corresponding path length by a statistical consideration, 
since we have sufficient knowledge of the dynamics of the system in the present 
(integrable) case. 

In the absence of interactions the path is composed 
of straight line segments. We denote the average length of these segments 
by $\langle l\rangle$. It can be calculated easily due to ergodicity:  
\begin{equation}\label{lav} 
\langle l\rangle =\frac{[(\pi_1^{(0)})^2+(\pi_2^{(0)})^2]^{1/2}}{\pi_1^{(0)}+\pi_2^{(0)}}
\equiv\frac{[2E_{int}]^{1/2}}{\pi_1^{(0)}+\pi_2^{(0)}}
\;\;, \end{equation} 
where $E_{int}$ denotes the conserved internal energy, cf. Eqs.\,(\ref{H2}) and (\ref{Cint}). 
Here we employed Eq.\,(\ref{phirelat}) and averaged the straight-line paths with the 
asymptotically uniform density over the square $[0,1[\times [0,1[$\,, choosing coordinates such that 
$\pi_{1,2}^{(0)}>0$.

Now, each increment of $I_1$ or $I_{21}$ (i.e. $I_2$) by one unit corresponds to the 
completion of one line segment. Due to the finite window size $\delta <1$, however, only a 
corresponding fraction of all incidents happening on both coordinate axes will be recorded on the average.    
Correcting for this, invoking ergodicity as before, 
the path length $L$ which leads to the measured 
incident numbers is simply obtained by:   
\begin{equation}\label{l}
L=\langle l\rangle\int_0^{I_1}\frac{\mbox{d}s}{\delta}
=\langle l\rangle\frac{I_1+I_{21}(I_1)}{\delta}  
\;\;, \end{equation}
where $\mbox{d}s$ is {\it along} the curve of the function $I_{21}(s)$. 
Practically, what appears on the right-hand side is the total registered 
incident number ($I\equiv I_1+I_2$) at a certain instant.    
  
In this way, we obtain a measure of the ``time'' interval $T$ during which    
$N$ incidents occurred: 
\begin{equation}\label{T}  
T\equiv L/[2E_{int}]^{1/2} 
\;\;, \end{equation} 
i.e., dividing the path length by the constant velocity.    

The point of this rather trivial example is that nowhere do we make use of 
the time parametrizing the evolution nor of the generally unknown path length.  
Rather, we derive $T$ from reparametrization-invariant localized measurements. 

Not surprisingly, the 
resulting ``time'' will not run smoothly, due to the  
coarse-grained description of the internal motion: as if we were reading an analog ``clock'' under 
a stroboscopic light. This is precisely the role of a Poincar\'e section 
with respect to the increasing invariant path length of an evolving trajectory. 


In order to illustrate the behavior of the ``time'' $T$, 
it is convenient to introduce the fictitious proper time (function): 
\begin{equation}\label{tau} 
\tau\equiv\int_0^t\mbox{d}t'\;\lambda (t')
\;\;. \end{equation}
Then, keeping the notation as simple as possible, 
with $x(\tau )\equiv x(t)$ for $x\in\{\vec q,\vec p,\phi_{1,2},\pi_{1,2}\}$, we obtain: 
$\partial_tx(t)=\lambda (t)\partial_\tau x(\tau )$. Applying this transformation 
to the definition of the canonical momenta and the equations of motion, 
the lapse function can be eliminated;      
this replaces $t$ by $\tau$ and $\lambda$ by 1 in 
Eqs.\,(\ref{p})--(\ref{pi2}) and Eqs.\,(\ref{Hq})--(\ref{Hphi2}), respectively. 
The resulting equations are reparametrization-invariant. 
Solutions of these equations are to     
be interpreted `physically' by introducing the inverse function $t(\tau )$ and 
$x(t)=x(\tau (t))$ \cite{Rovelli90}.

Integrating the free internal motion with respect to the proper time, which replaces the 
integral in Eq.\,(\ref{phifree}) by $\tau$, we show $T$ as a function of 
$\tau$ in \underline{Fig.}\,2a) and \underline{Fig.}\,2b). For the two runs differing in the 
total (computer) time $\tau$, we find that after a short while, i.e. already at low 
incident numbers, the constructed physical ``time'' $T$ approximates qualitatively well the 
proper time $\tau$. The fluctuations on top of the observed linear dependence 
naturally reflect the stochastic internal motion. The fact that the slopes are 
consistently smaller than one can be attributed to the bias towards longer-than-average 
pieces of trajectory, 
which is introduced by measuring the incidents close to the origin and extrapolating 
from there, see Eqs.\,(\ref{lav}) and (\ref{l}).   
  
Finally, employing the 
reparametrization-invariant ``time'', which runs approximately parallel to 
the unphysical proper time, $T\approx\kappa\tau$ ($\kappa$ constant), we succeed to 
describe the external 
motion with respect to the physical internal ``clock'' constructed here. 
Recalling that the external motion is given by Eq.\,(\ref{qevol}), we obtain: 
$\vec q_\parallel(\tau)=\vec q_\parallel^{\;(0)}+\tau\vec p\approx
\vec q_\parallel^{\;(0)}+\kappa^{-1}T\vec p$.  

Interestingly, the coarse-grained `jumpiness' of $T$ introduces a corresponding 
stochastic component into the external motion. This would be recognized, if  
global relational data or results of increasingly fine-grained local measurements were available, 
with which to compare the change of $\vec q_\parallel$. We will further study the consequences 
of this stroboscopic effect in Section\,4\,.   

\subsection{Quasi-local analysis of ``time is change'' with interactions}
Now we consider the case of the internal motion on the torus with the interactions 
($\propto\omega^2,\Omega^2$) turned on. The interactions 
additionally mix the trajectories in phase space, since the potentials are of  
inverted oscillator type. Slightly simplifying the ensuing calculations, we set $\omega^2=\Omega^2$ henceforth. 
This leaves the motion parallel to the diagonal $\widetilde\phi_1=\widetilde\phi_2$ exponentially unstable, while 
the orthogonal motion is free. A typical trajectory for the case of still relatively 
weak interaction is shown in \underline{Fig.}\,3\,.
  
The microstates of the internal part of the system can be described by the phase space 
coordinates $x\equiv (\phi_1,\phi_2;\pi_1,\pi_2)$. However, in general, we will not be able to 
follow the deterministic evolution through sequences of microstates for any sufficiently 
complex nonlinear system. Therefore, we develop a coarse-graining statistical approach based 
on probability distributions. Let  
$P(x)\mbox{d}x$ denote the expected number of microstates in the volume element $\mbox{d}x$ 
at $x$. Then, we may characterize macrostates of the system by giving their 
coordinates $Q^i,\;i=1,\dots ,n$ in the $n$-dimensional state space: 
\begin{equation}\label{Qc} 
Q^i\equiv\langle {\cal Q}^i\rangle\equiv\int\mbox{d}x\;P(x){\cal Q}^i(x)
\;\;, \end{equation} 
with $\{{\cal Q}^i\}$ denoting the relevant set of observables. It is assumed that all information necessary to answer our particular questions about the system is encoded in these 
observables \cite{Ariel01}.   
   
The conserved internal energy, see Eq.\,(\ref{Cint}), presents an important observable: 
\begin{equation}\label{Hint}      
{\cal H}_{int}(x)\equiv \pi_1^{\;2}+\pi_2^{\;2}
-\omega^2(\widetilde\phi_1+\widetilde\phi_2)^2
\;\;, \end{equation}
where we absorbed an inessential factor 2 into the definition for convenience. 
    
Then, the `prior' distribution $P_c(x|H_{int})$ which best reflects our 
mostly lacking information about the state of the system, 
given the conserved energy, is obtained by maximizing the 
entropy: 
\begin{equation}\label{entropy}
S[P]\equiv -\int\mbox{d}x\;P(x)\ln P(x)
\;\;, \end{equation} 
subject to the constraint $\langle {\cal H}_{int}\rangle =H_{int}$, with this constant being  
fixed by the initial conditions. 
The result is the canonical distribution: 
\begin{equation}\label{Pc}
P_c(x|H_{int})=Z_c^{-1}e^{-\beta {\cal H}_{int}(x)} 
\;\;, \end{equation} 
with the partition function $Z_c$ and Lagrange multiplier $\beta$, respectively, to be   
calculated from: 
\begin{equation}\label{Zbeta} 
Z_c\equiv\int\mbox{d}x\;e^{-\beta {\cal H}_{int}(x)}\;\;, \;\;\;
H_{int}=-\frac{\partial}{\partial\beta}\ln Z_c
\;\;, \end{equation}
where $\int\mbox{d}x\equiv\int_{-\infty}^\infty\mbox{d}\pi_1\int_{-\infty}^\infty\mbox{d}\pi_2
\int_0^1\mbox{d}\phi_1\int_0^1\mbox{d}\phi_1$. Thus we recover expressions which are familiar from statistical mechanics.  

Straightforward calculation yields the partition function:   
\begin{equation}\label{Zconf}
Z_c=\frac{\pi}{\beta}Z_{c,conf}=\frac{\pi}{\beta}J(1,1)
\;\;, \end{equation} 
where $Z_{c,conf}$ denotes its configurational factor.  
It is given here in terms of the integral:
\begin{eqnarray}\label{Jintegral} 
J(a,b)&\equiv&\frac{1}{\beta\omega^2}\int_0^{a\sqrt{\beta\omega^2}}\mbox{d}s_1 
\int_0^{b\sqrt{\beta\omega^2}}\mbox{d}s_2\;
e^{(s_1+s_2)^2} 
\\ [1ex] 
&=&\frac{1}{2}\sqrt{\frac{\pi}{\beta\omega^2}}\left (
(a+b)\;\mbox{erfi}((a+b)\sqrt{\beta\omega^2})
-a\;\mbox{erfi}(a\sqrt{\beta\omega^2})
-b\;\mbox{erfi}(b\sqrt{\beta\omega^2})\right ) 
\nonumber \\ [1ex] \label{Jintegral1}
&\;&-\frac{1}{2\beta\omega^2}\left (e^{(a+b)^2\beta\omega^2}-e^{a^2\beta\omega^2}
-e^{b^2\beta\omega^2}+1\right )
\\ [1ex] 
&=&ab\left (1+\beta\omega^2[\frac{1}{3}a^2+\frac{1}{2}ab+\frac{1}{3}b^2]\right . 
\nonumber \\ [1ex] \label{Jintegral2} 
&\;&\;\;\;\;\;\left .+\frac{1}{2}(\beta\omega^2)^2[\frac{1}{5}a^4+\frac{1}{2}a^3b+\frac{2}{3}a^2b^2
+\frac{1}{2}ab^3+\frac{1}{5}b^4]+\mbox{O}[(\beta\omega^2)^3]\right )
\;\;, \end{eqnarray} 
involving the imaginary error function,  
$\mbox{erfi}(x)\equiv 2\pi^{-1/2}\int_0^x\mbox{d}s\;\exp (s^2)$. We will make 
more use of this integral in the following. 


For example, incorporating the small-$\beta$ expansion, we calculate the relation between $H_{int}$ and $\beta$:
\begin{equation}\label{Eintbeta}
\frac{H_{int}}{\omega^2}=\frac{1}{\beta\omega^2}-\frac{7}{6}-\frac{127}{180}\beta\omega^2
+\mbox{O}[(\beta\omega^2)^2]
\;\;, \end{equation} 
using Eqs.\,(\ref{Zbeta}), (\ref{Zconf}), and (\ref{Jintegral2}). 
Here the error in comparison with the exact result rapidly 
decreases with energy and is less than $5\%$ for $H_{int}/\omega^2>1$.                                 
We are presently interested in the positive energy regime, since only 
there the trajectories can explore all of the torus surface, cf. Eq.\,(\ref{Hint}).

In order to improve the prior distribution $P_c$  
with the help of quasi-local measurements, we again consider simple observables for 
illustration. We define two ``window functions'': 
\begin{equation}\label{incfct} 
{\cal I}_1(x)\equiv\Theta (\phi_1-\epsilon )\Theta (\delta -\phi_1)\Theta (\phi_2)\Theta (\epsilon -\phi_2) 
\;\;, \end{equation} 
with $0<\epsilon\ll\delta <1$; ${\cal I}_2$ is defined by the analogous expression 
with $\phi_1$ and $\phi_2$ exchanged. Thus ${\cal I}_1$ and ${\cal I}_2$ project out 
small rectangles along the $\phi_1$- and $\phi_2$-axis, respectively, which do not overlap. 
Here we explicitly introduced the small but finite window width $\epsilon$, which in any case 
is necessary for a correct counting of incidents in numerical simulations with a finite resolution. In the following we will adapt to the present case the measurement of incident numbers.     

We decompose the prior distribution with respect to the windows determined by ${\cal I}_1$ 
and ${\cal I}_2$: 
\begin{eqnarray}
P_c(x|H_{int})&=&[{\cal I}_1(x)+{\cal I}_2(x)]P_c(x|H_{int})+[1-{\cal I}_1(x)-{\cal I}_2(x)]P_c(x|H_{int})
\nonumber \\ [1ex]
\label{decomp}
&\equiv&P_w'(x|H_{int})+\bar P_w(x|H_{int})
\;\;. \end{eqnarray}
While $\bar P_w$ describes their complement, the distribution $P_w'$ is the one which is related  
to measurements within the  
windows. Its normalized counterpart $P_w$ is:  
\begin{equation}\label{reducedprior} 
P_w(x|H_{int})\equiv\frac{[{\cal I}_1(x)+{\cal I}_2(x)]P_c(x|H_{int})}
{\int\mbox{d}x\;[{\cal I}_1(x)+{\cal I}_2(x)]P_c(x|H_{int})}
=\frac{\beta}{2\pi}\frac{[{\cal I}_1(x)+{\cal I}_2(x)]e^{-\beta {\cal H}_{int}(x)}}
{J(\delta ,\epsilon )-J(\epsilon ,\epsilon )}
\;\;, \end{equation}  
employing the configurational integral of Eq.\,(\ref{Jintegral}). 
  
Then, the distribution $P_w(x|H_{int};\langle {\cal I}_i\rangle)$ which best reflects the information contained in  
the prior distribution $P_w(x|H_{int})$ and in the data from measurements of the 
window functions is  
obtained by maximizing the entropy \cite{Ariel01}: 
\begin{equation}\label{entropy1}
S[P]\equiv -\int\mbox{d}x\;[{\cal I}_1(x)+{\cal I}_2(x)]P(x)\ln \frac{P(x)}{P_w(x|H_{int})}
\;\;, \end{equation} 
subject to the constraints: 
\begin{equation}\label{Iconstraints}
\langle {\cal I}_i\rangle =\frac{I_i}{I_1+I_2}\;|_{i=1,2}
\;\;. \end{equation}  
Here we determine the average of the incident functions, i.e. the (total)  
probabilities of observing an incident in the respective windows, in terms of the 
measured incident numbers. This obviously presents a crude coarse-graining.  
An improved description is obtained, for example, by binning the incident 
numbers with respect to the main axis of each window. Many, more detailed measurements 
can be envisaged, but the simplest ones will suffice here.  
 
It is straightforward to show that this procedure yields a grand-canonical 
distribution:
\begin{equation}\label{P}
P_w(x|H_{int};\langle {\cal I}_i\rangle )=Z^{-1}[{\cal I}_1(x)+{\cal I}_2(x)]
e^{-\beta {\cal H}_{int}(x)-\alpha_1{\cal I}_1(x)-\alpha_2{\cal I}_2(x)}
\;\;, \end{equation} 
where, in this case, the partition function and Lagrange multipliers are determined by:  
\begin{eqnarray}Z&\equiv&\int\mbox{d}x\;[{\cal I}_1(x)+{\cal I}_2(x)]
e^{-\beta {\cal H}_{int}(x)-\alpha_1{\cal I}_1(x)-\alpha_2{\cal I}_2(x)}
\nonumber \\ [1ex] 
\label{Z}
&=&\int\mbox{d}x\;[{\cal I}_1(x)\lambda_1+{\cal I}_2(x)\lambda_2]
e^{-\beta {\cal H}_{int}(x)} 
\;\;, \end{eqnarray}
introducing the fugacities $\lambda_i\equiv\exp (-\alpha_i)\;|_{i=1,2}$, and: 
\begin{equation}\label{fugacity}
\langle {\cal I}_i\rangle =-\frac{\partial}{\partial\alpha_i}\ln Z 
=\lambda_i\frac{\partial}{\partial\lambda_i}\ln Z\;|_{i=1,2}
\;\;, \end{equation} 
together with the constraints (\ref{Iconstraints}). Here   
$\beta$ is considered to be a known feature of the prior distribution $P_c$, previously determined in  
Eq.\,(\ref{Eintbeta}). 
 
The present situation differs in an important way 
from the usual one in statistical mechanics, where $H_{int}$ and $\langle {\cal I}_i\rangle$ would {\it all} correspond to conserved quantities and be treated on an equal footing. 
Presently, the 
(window functions related to the) incident numbers are evolving 
quantities which we measure, in order to learn about the change occurring in   
the system.    
 
The grand-canonical partition function can be calculated directly, resulting in:
\begin{equation}\label{Zgrand}
Z=\frac{\pi}{\beta}Z_{conf}=\frac{\pi}{\beta}(\lambda_1+\lambda_2)
[J(\delta ,\epsilon )-J(\epsilon ,\epsilon )] 
\;\;, \end{equation} 
where $Z_{conf}$ is the configurational factor of this partition function. Furthermore, we obtain: 
\begin{equation}\label{lambda} 
\langle {\cal I}_i\rangle =\frac{\lambda_i}{\lambda_1+\lambda_2}\;|_{i=1,2} 
\;\;, \end{equation} 
which implies $\lambda_1/\lambda_2=\langle {\cal I}_1\rangle /\langle {\cal I}_2\rangle$. 
We set $\lambda_i=C\langle {\cal I}_i\rangle\;|_{i=1,2}$, with a 
common (undetermined) constant $C$. Incorporating these results, the distribution follows:   
\begin{equation}\label{Pw}
P_w(x|H_{int};\langle {\cal I}_i\rangle )=\frac{\beta}{\pi}
\frac{[\langle {\cal I}_1\rangle {\cal I}_1(x)+\langle {\cal I}_2\rangle {\cal I}_2(x)]}
{J(\delta ,\epsilon )-J(\epsilon ,\epsilon )}e^{-\beta {\cal H}_{int}(x)} 
\;\;, \end{equation}
using $\langle {\cal I}_1+{\cal I}_2\rangle =1$ and cancelling the common factor $C$. 
  
Finally, re-normalizing $P_w(x|H_{int};\langle {\cal I}_i\rangle )$ and using the resulting 
distribution in place of $P_w'$, cf. Eq.\,(\ref{decomp}), we obtain the properly 
normalized distribution for the whole phase space: 
\begin{equation}\label{Pfinal}
P(x|H_{int};\langle {\cal I}_i\rangle )=2[\langle {\cal I}_1\rangle {\cal I}_1(x)
+\langle {\cal I}_2\rangle {\cal I}_2(x)]
P_c(x|H_{int})+[1-{\cal I}_1(x)-{\cal I}_2(x)]P_c(x|H_{int})
\;\;, \end{equation} 
which is updated by measuring the incident numbers $I_{1,2}$ and  
employing Eq.\,(\ref{Iconstraints}).  
 
With the distribution at hand, we could proceed similarly as in the previous 
Section\,3.1\,, trying to estimate the average path length related to the increasing 
incident numbers in particular, in order to gain a measure of the 
change taking place in the system. 

However, in the following we proceed differently, in a way which appears more suitable 
to further generalization. We introduce the 
Fisher-Rao information metric for the purpose of quantifying the 
change due to the chaotic, even if deterministic motion from one 
configuration to the next \cite{Fisher,Rao}. It is the  
uniquely determined Riemannian metric (except for an overall multiplicative constant) 
on the space of states which are probability distributions \cite{Ariel01}. In 
our present case the states are simply described by the pair of    
coordinates $Q^i\equiv \langle {\cal I}_i\rangle \;|_{i=1,2}\in [0,1]$, considering the coordinate $H_{int}$ 
to be fixed at a constant value. Then, the `distance' $\mbox{d}s$ 
between the states $P(x|H_{int};Q^k)$ and 
$P(x|H_{int};Q^k+\mbox{d}Q^k)$ is       
given by: 
\begin{equation}\label{FisherRao}
\mbox{d}s^2=g_{ij}\mbox{d}Q^i\mbox{d}Q^j  
\;\;, \end{equation}  
with the metric: 
\begin{eqnarray}\label{metric1}
g_{ij}&=&\int\mbox{d}x\;P(x|H_{int};Q^k)\;
\frac{\partial\ln P(x|H_{int};Q^k)}{\partial Q^i}  
\frac{\partial\ln P(x|H_{int};Q^k)}{\partial Q^j} 
\\ [2ex] 
\label{metric2} 
&=&2\frac{J(\delta ,\epsilon )-J(\epsilon ,\epsilon )}{J(1,1)}(Q^i)^{-1}\delta_{ij}
\;\;, \end{eqnarray}
employing Eqs.\,(\ref{Pc}), (\ref{Zconf}), (\ref{Jintegral}), and particularly Eq.\,(\ref{Pfinal}) 
in the explicit calculation. Note that the coordinates, being probabilities, are constrained by 
$Q^1+Q^2=1$, which further simplifies the result obtained here.  
  
However, we would like to express Eqs.\,(\ref{FisherRao})--(\ref{metric2}) in terms of the directly measured 
incident numbers. Thus, with the help of Eqs.\,(\ref{Iconstraints}), we obtain: 
\begin{eqnarray}\label{deltas1}
\Delta s&=&2\frac{J(\delta ,\epsilon )-J(\epsilon ,\epsilon )}{J(1,1)}
\frac{|I_2\Delta I_1-I_1\Delta I_2|}{\sqrt{I_1I_2}(I_1+I_2)}
\\ [1ex] \label{deltas2}
&=&2\frac{J(\delta ,\epsilon )-J(\epsilon ,\epsilon )}{J(1,1)(I_1+I_2)}
\left (\sqrt{\frac{I_2}{I_1}}\Delta I_1+\sqrt{\frac{I_1}{I_2}}\Delta I_2\right )
\;\;, \end{eqnarray} 
where the second equality is due to the dichotomous character of the 
pair of increments $\Delta I_1$ and $\Delta I_2$; i.e., we have $\Delta I_k=0,1$, however, the two 
increments present the results of two mutually exclusive measurements which cannot  
happen in coincidence.   
  
Several remarks are in order here. 
First, the Eq.\,(\ref{deltas2}) is invariant under the similarity transformation 
$I_k\rightarrow\xi\cdot I_k\;|_{k=1,2}$ and $\Delta s\rightarrow\Delta s/\xi$. 
Thus, generally, 
the amount of ``change per incident" decreases inversely with the total number of incidents, i.e., loosely speaking, with the age of the evolution going on. This seems natural in a statistical context, but less familiar in the context of dynamics, where we are accustomed to a linearly flowing time. 

Second, the dichotomous increments of the incident numbers can be mapped on a binary sequence of zeros and ones, 
for example, according to $(\Delta I_1=1)\rightarrow 0$ and $(\Delta I_2=1)\rightarrow 1$. Then, the number 
of zeros and ones in the produced bitstring is simply related to $I_1-1$ and $I_2-1$, respectively. 
Thus, one could compare the behavior of the function $\Delta s$ with corresponding ones 
for different dynamical models or with randomly generated sequences.  

Third, the statistical factors in the results of Eqs.\,(\ref{metric2})--(\ref{deltas1}), 
encoding the dynamics, are nicely factorized from 
the quantities incorporating the evolving observables $I_1$ and $I_2$; an 
analogous factorization is seen in Eq.\,(\ref{T}).  
We believe that the very simple structure of our results is indeed of a more general kind and similarly will 
be encountered, whenever one succeeds to identify pairs of dichotomous observables, which somehow 
reflect the symmetry of the system. 

Furthermore, the 
generalization involving n-tuples of exclusive observables can be worked out along these lines.      

Coming back to the construction of a reparametrization invariant ``time'' 
$T$, we now set: 
\begin{equation}\label{Ts} 
\Delta T\equiv\frac{N\Delta s}{\sqrt{H_{int}}}  
\;\;, \end{equation} 
with $N\equiv I_1+I_2$, i.e. stretching the time with an extra factor $N$, in order 
to compensate for the aging of the measure $\Delta s$, which we mentioned.   
Furthermore, we divide out the square-root-of-energy factor, which is determined 
by the initial conditions. This results in a scaling with energy, similarly as in Section\,3.1\,. At high energy $H_{int}$, Eq.\,(\ref{Hint}), equals (two times) the kinetic energy, which 
might seem more appropriate here. However, it is by itself not a  
reparametrization invariant quantity in the present interacting case. 
  
We calculate the ``time'' passed by accumulating the time steps $\Delta T$. The constant, though energy 
dependent, overall prefactor contained in the result of Eq.\,(\ref{Ts}) is set equal to one, recalling 
that the Fisher-Rao metric which enters here, see Eqs.\,(\ref{FisherRao})--(\ref{metric2}), is unique only 
up to a constant factor. The \underline{Fig.}\,4 shows the resulting 
behavior of the parametrization invariant ``time'' $T$ as a function of the proper time $\tau$, 
Eq.\,(\ref{tau}), which is also incorporated in the present numerical simulations. 
For the two initial conditions, differing in the initial velocities, we observe the 
expected scaling with energy. Furthermore, while the size of individual fluctuations is comparable to what was found in the noninteracting case 
of \underline{Fig.}\,2, we notice here characteristic deviations from   
the average linear dependence over longer proper time 
scales. Such nonlinear behavior becomes even more prominent in \underline{Fig.}\,5, with 
long time excursions. 

In any case, this demonstrates the existence of a monotonous ``time'' 
function also for the present more strongly irregular system, 
as compared to the ergodic (but integrable) one of Section\,3.1\,. 
Presumably, an approximately linear dependence will only become recognizable for much 
longer runs, i.e. for larger incident numbers. On the other hand,  
of course, the statistical construction of $T$ based on quasi-local 
observables could be refined by considering a higher-dimensional state space 
from the outset, cf. Eq.\,(\ref{Qc}), i.e. incorporating more observables.            
  
This completes our discussion of two examples of the construction of a coarse-grained 
parametrization invariant evolution parameter. 

\section{Stroboscopic quantization?} 
It seems worth while to draw attention to a possible connection between 
the present subject and the {\it deterministic classical systems which are effectively quantum 
mechanical}, as  
recently discussed by 't\,Hooft, see Ref.\,\cite{tHooft01} and further 
references therein. 

The total number of incidents obtained from our exclusive observables always increases 
by one. These are the ``ticks of the clock'', numbered by $n\in\mathbf{N}$. However, 
via the constructed ``time'' sequence $\{T_n\}$, the proper times $\tau (T_n)$ vary  
irregularly, in general, due to the quasi-periodic internal motion. Therefore, the free 
external motion, 
$\vec q_\parallel(T)\equiv\vec q_\parallel(\tau (T))=\vec q_\parallel^{\;(0)}+\tau (T)\vec p$, 
shows stochastic behavior in terms of $T$. We emphasize that $\tau (T)$ here is {\it not}   
directly related to $\tau (t)$ introduced in Eq.\,(\ref{tau}).    
 
Let us assume that we could exhaust all possible {\it invariant quasi-local observables} 
and cannot 
further resolve the ``time between the incidents'' marking $T$ and $T+\delta T$. 

That there arises a discreteness in the optimal ``time'' sequence generated in this way in a generic 
nonintegrable system \cite{Schuster} seems plausible for two 
reasons. First, by realistically restricting the observables to be localized near to the observer, we loose information 
about the full trajectories in coordinate space. Second, by insisting on gauge invariant observables, we loose the possibility to make use of additional momentum space observables 
(except for a possible limited number of constants of motion), or higher proper time 
derivatives, in order to 
reconstruct the full trajectories and a smooth evolution parameter based on them.  

Anyhow, for given initial conditions, the incident counts as well as the coordinates 
$\vec q_\parallel(T_n)$ present discrete physical 
data which are deterministically determined in our model. 
(The constant orthogonal components are of no interest here.) 
Then, the question arises, can we still give an invariant meaning to 
and predict the future of the free external motion,   
instead of letting the system evolve and measuring? - In general, the answer is `No', since neither 
the future discrete values of $T$, see Eq.\,(\ref{Ts}), for example, nor the value of 
$\tau$ at those future 
instants will be predictable.   

However, motivated by the two practical examples of Section\,3\,, we introduce 
the notion of a well-behaved ``time'', which will be sufficient to maintain some predictability 
indeed. 

Let a {\it well-behaved ``time''} $T$ be such that the onedimensional motion of a 
free particle, co-ordinated in a reparametrization-invariant way by the sequence $\{ q(T_n)\}$, 
is of limited variation: 
\begin{eqnarray}
&\;&\{T_n\}\;\;\mbox{is well-behaved} 
\nonumber \\ [1ex] 
&\;&\Longleftrightarrow\;\;\exists\;\bar\tau ,\Delta\tau:\;
\bar\tau n-\Delta\tau\leq\tau (T_n)\leq\bar\tau n+\Delta\tau\;,\;\;\forall n\in{\mathbf N}
\nonumber \\ [1ex] \label{perfect2}
&\;&\Longleftrightarrow\;\;\exists\;\;\bar q,\Delta q:\;
\bar qn-\Delta q+q^{(0)}\leq q(T_n)\leq\bar qn+\Delta q+q^{(0)}\;,\;\;\forall n\in{\mathbf N}
\;\;, \end{eqnarray} 
with $\bar q\equiv\bar\tau p^{(0)},\;\Delta q\equiv\Delta\tau p^{(0)}$, and where $q^{(0)},p^{(0)}$ 
denote the initial data.  
This is the next-best we can expect from a reasonable ``time'', motivated by Newtonian physics here. 
It is important to realize that there is no hope for continuity or periodicity, since our 
construction of $T$ is based generally on the ergodicity of a quasi-periodic process and related Poincar\'e sections.  
  
Assuming a well-behaved ``time'', we see that the sequence of points $\{ q(T_n)\}$   
can be mapped into a regular lattice 
of possibly overlapping cells of size $2\Delta q$ and spacing $\bar q$. 
``As the clock ticks'', i.e. $n\rightarrow n+1$, the particle moves from one cell 
to the next. On the average this takes a proper time $\bar\tau$ and physical ``time'' 
$\bar T$. Here we assume for convenience that $\{ T_n\}$ itself is of limited variation, 
$\exists\;\bar T,\Delta T:T_n=\bar Tn\pm\Delta T,\;\forall n\in{\mathbf N}$. 
In the simplest example, in Section\,3.1, we 
have $\bar T=\delta^{-1}(\pi_1^{(0)}+\pi_2^{(0)})^{-1}$ and $\Delta T=0$, showing a 
dependence on the initial momenta and the scale of localization of the observed incidents.  


Thus, for a well-behaved ``time'' sequence (of limited variation), we have  
mappings from ``clock ticks'' to (``time'' as well as) space intervals. We identify the space 
intervals, containing the respective $q(T_n)$, with 
{\it primordial states}:
\begin{equation}\label{states}
|n)\equiv [\bar qn-\Delta q+q^{(0)},\bar qn+\Delta q+q^{(0)}] 
\;\;. \end{equation} 
In order to handle the subtle limit of an infinite system, we impose reflecting boundary 
conditions and distinguish left- and right-moving states. Finally, then, the  
evolution is described by the deterministic rules: 
\begin{eqnarray}\label{R1}
n&\longrightarrow&n+1\;,\;\;n\in{\mathbf N}  
\;\;, \\ [1ex]
\label{R2}
|n)&\longrightarrow&|n+1)\;,\;\;-N+1\leq n\leq N-1
\;\;, \\ [1ex]
\label{R3}
|N)&\longrightarrow&|-N+1)
\;\;, \end{eqnarray} 
with $2N$ states in all, $|-N+1),\;\dots\;,|N)$; states with a negative (positive) label 
correspond to left- (right-)moving states, according to negative (positive) particle momentum.  
Reaching the states $|0)$ or $|N)$, the particle changes its direction of motion. 

The rules (\ref{R1})--(\ref{R3}) can altogether be represented by a unitary evolution operator:   
\begin{equation}\label{U}
U(\delta T =\bar T)\equiv e^{-iH\bar T}=e^{-i\pi (N+1/2)/N}\cdot
\left (\begin{array}{ccccc}
0&\;&\;&\;&1 \\
1&0&\;&\;&\; \\
\;&1&0&\;&\; \\
\;&\;&\ddots&\ddots&\; \\
\;&\;&\;&1&0 
\end{array}\right )
\;\;, \end{equation}
which acts on the $2N$-dimensional vector composed of the primordial states; 
the overall phase factor is introduced for later convenience. Here  
$\bar T$ is the natural scale for the Hamiltonian $H$. 

In the following analysis we apply 't\,Hooft's 
method, who considered the discrete and strictly periodic motion of a classical 
particle on a circle, introducing a Hilbert space based on the primordial states \cite{tHooft01}.  
The evolution operator turns out to be diagonal with respect to the discrete Fourier 
transforms of the states $|n)$. We define the basis functions: 
\begin{equation}\label{kBasis} 
\langle k|n)\equiv f_k(n)\equiv (2N)^{-1/2}\exp (\frac{i\pi kn}{N})   
\;\;. \end{equation}
They present a complete orthonormal basis, since $\sum_{k=-N+1}^N (n'|k\rangle\langle k|n)=\delta_{n'n}$, 
with $(n|k\rangle\equiv\langle k|n)^\ast$, and noting that $f_k(n)=f_n(k)$. 
This yields indeed: 
\begin{equation}\label{Udiag}
\langle k'|U|k\rangle =\delta_{k'k}\exp (-i\pi [N+\frac{1}{2}-k]/N) 
\;\Leftrightarrow\;\langle m'|U|m\rangle =\delta_{m'm}\exp (-\frac{2\pi i}{2s+1}[s+m+\frac{1}{2}]) 
\;\;, \end{equation}
where the equivalence follows from $2s+1\equiv 2N$ and relabeling 
and replacing the states $|k\rangle$ by states $|m\rangle$, with $m\equiv -k+1/2$ and $-s\leq m\leq s$. 
The phase factor of Eq.\,(\ref{U}) contributes the additional terms which assure a positive definite (bounded) spectrum similar to the harmonic oscillator case.   

Recalling the algebra of $SU(2)$ generators, and with $S_z|m\rangle =m|m\rangle$ 
in particular, we obtain the Hamiltonian: 
\begin{equation}\label{Hdiag} 
H=\frac{2\pi}{(2s+1)\bar T}(S_z+s+\frac{1}{2}) 
\;\;, \end{equation}
i.e., diagonal with respect to $|m\rangle$-states of the half-integer representations determined by $s$. 

Continuing with standard notation, we have $S^2\equiv S_x^{\;2}+S_y^{\;2}+S_z^{\;2}=s(s+1)$, which suffices 
to obtain the following identity: 
\begin{equation}\label{Hsq} 
H=\frac{2\pi}{(2s+1)^2\bar T}(S_x^{\;2}+S_y^{\;2}+\frac{1}{4})+\frac{\bar T}{2\pi}H^2 
\;\;. \end{equation} 
Furthermore, using $S_\pm\equiv S_x\pm iS_y$, we introduce coordinate and conjugate 
momentum operators: 
\begin{equation}\label{qp}
q\equiv\frac{1}{2}(aS_-+a^\ast S_+)\;\;,\;\;\; 
p\equiv\frac{1}{2}(bS_-+b^\ast S_+)
\;\;, \end{equation} 
where $a$ and $b$ are complex coefficients. Calculating the basic commutator with the help 
of $[S_+,S_-]=2S_z$ and using Eq.\,(\ref{Hdiag}), we obtain: 
\begin{equation}\label{commutator} 
[q,p]=i(1-\frac{\bar T}{\pi}H)
\;\;, \end{equation} 
provided we choose $\Im (a^\ast b)\equiv -2(2s+1)^{-1}$. Incorporating this choice, 
we calculate: 
\begin{equation}\label{sumsqu}
S_x^{\;2}+S_y^{\;2}=\frac{(2s+1)^2}{4}\left (
|a|^2p^2+|b|^2q^2-(\Im a\cdot\Im b+\Re a\cdot\Re b)\{ q,p\}\right )
\;\;. \end{equation}  
In order to obtain a reasonably simple Hamiltonian in the infinite system limit ($s\rightarrow\infty$), 
we finally choose: 
\begin{equation}\label{ab} 
a\equiv i\sqrt{\frac{\bar T}{\pi}}\;\;,\;\;\;b\equiv\frac{2}{2s+1}\sqrt{\frac{\pi}{\bar T}}
\;\;. \end{equation}  
Then, defining $\omega\equiv 2\pi /(2s+1)\bar T$, the previous Eq.\,(\ref{Hsq}) becomes: 
\begin{equation}\label{Hsq1} 
H=\frac{1}{2}p^2+\frac{1}{2}\omega^2q^2+\frac{\bar T}{2\pi}(\frac{1}{4}\omega^2+H^2)
\;\;, \end{equation}
showing a nonlinearly modified harmonic oscillator Hamiltonian at this stage, 
which corresponds to the result of Ref.\,\cite{tHooft01}. 
  
Following our construction of the reparametrization-invariant ``time'', 
we presently keep $\bar T$ finite 
while considering the infinite system limit. 
Thus, for $s\rightarrow\infty$, we have $\omega\rightarrow 0$ and obtain: 
\begin{equation}\label{Hpsq}
H=\frac{\pi}{\bar T}\left (1-(1-\frac{\bar T}{\pi}p^2)^{1/2}\right ) 
\;\;, \end{equation} 
which has the low-energy limit $H\approx p^2/2$. On the other side,  
the energy is bounded from above 
by $\pi /\bar T$, since we have $(\bar T/\pi )p^2=4(2s+1)^{-2}S_x^{\;2}$ 
($\leq 1$, for $s\rightarrow\infty$, when diagonalized). 

Interestingly, towards the upper bound the violation of the basic 
quantum mechanical commutation relation, as calculated in Eq.\,(\ref{commutator}),    
becomes maximal, i.e., there $q$ and $p$ commute and behave like classical observables. 

We conclude here with calculating the matrix elements of the operators $q$ and $p$ between 
primordial states. Relating the primordial states $|n)$ to the $|m\rangle$-states with the 
help of Eq.\,(\ref{kBasis}) and using $S_\pm |m\rangle =[s(s+1)-m(m\pm 1)]^{1/2}|m\pm 1\rangle$, 
we obtain: 
\begin{eqnarray}\label{Spm} 
(n'|S_-\pm S_+|n)&=&\frac{1}{2s+1}
\left (\exp (\textstyle{-\frac{2\pi i}{2s+1}n})
\pm\exp (\textstyle{\frac{2\pi i}{2s+1}n'})\right )
\nonumber \\ [1ex] 
&\;& 
\cdot\sum_{m=-s}^{s}\exp (\textstyle{\frac{2\pi i}{2s+1}(n'-n)(m-1/2)})
[s(s+1)-m(m+1)]^{1/2}
\;\;. \end{eqnarray} 
For large $s$ the summation is replaced by an integral. Then, using Eqs.\,(\ref{qp}) and 
(\ref{ab}), the results are ($s\rightarrow\infty$): 
\begin{eqnarray}\label{qmatrix} 
(n'|q|n)&=&\frac{1}{2}\sqrt{\pi\bar T}\frac{J_1(\pi (n'-n))}{n'-n}\frac{n'+n}{2} 
\;\;, \\ [2ex]  
\label{pmatrix} 
(n'|p|n)&=&\frac{1}{2}\sqrt{\frac{\bar T}{\pi}}\frac{J_1(\pi (n'-n))}{n'-n} 
\;\;, \end{eqnarray} 
where $J_1$ denotes an ordinary Bessel function of the first kind. Thus, neither the position nor 
the momentum operator  
is diagonal in the basis of the primordial states. 
  
In any case, we find here once more an emergent quantum model based on a 
deterministic classical evolution, similar as in Ref.\,\cite{tHooft01}. 

The interesting feature presently is 
that the construction of a reparametrization-invariant ``time''  
based on quasi-local observables naturally induces the particular stochastic 
features in the behavior of the external particle motion. The remaining predictable 
aspects of its motion can then most simply be described by a unitary 
discrete-time quantum mechanical evolution.      
  
\section{Conclusions}    
Our construction of a reparametrization-invariant ``time'' is motivated by the observation 
that ``time passes'' 
when there is an observable change, which is localized with the observer. More precisely, 
necessary are incidents, i.e. observable unit changes, which are recorded, and from which invariant quantities characterizing the change of the evolving system can be derived.    
   
A basic ingredient is the assumption of ergodicity, such that the system explores 
dynamically the whole allowed energy surface in phase space. Generally, then there 
will be sufficiently frequent usable incidents next to the observer. 

We illustrate this in Section\,3 with a simple timeless model of a nonrelativistic particle 
moving classically in five dimensions, of which two internal ones are compactified to form 
a torus. Defining suitable window functions, localized observations of incidents correspond 
here to the trajectory passing by the window. Thus, the incidents reflect properties 
of the dynamics with respect to (subsets of) Poincar\'e sections. Roughly, the 
passing ``time'' corresponds to the observable change there.      
   
In the first example in Section\,3.1 the dynamics is very simple and even  
integrable \cite{Schuster}. We know that the system is ergodic in this case and use this 
fact, in order to calculate statistically, based on quasi-local incident counts, the invariant 
path length run by the system.  
This provides an invariant measure of ``time''. We find that the proper time is 
linearly related to this, however, subject to stochastic fluctuations.       
  
In the second example, Section\,3.2, we have additional interactions which mix 
the trajectories in a much more complicated way. We assume that the system is 
ergodic. Our statistical formalism then allows to measure the 
change of the phase space distribution which is updated by the information 
coming from the incident counts. Essential ingredient is the Fisher-Rao 
information metric \cite{Ariel01,Fisher,Rao}. This calculation again provides 
an invariant ``time'' function, which behaves qualitatively similar as in 
the previous example, however, with characteristic long-time deviations 
from an average linear relation to proper time.    
  
It will be interesting to generalize the present model  
to relativistic systems, as well as to generalize the statistical formalism. 
A related question is, which are the (discrete) limitations of such a construction 
of ``time'' based on localized observables, i.e. how close can one get to a linearly flowing Newtonian time, in a generic nonintegrable system. 
    
Finally, we pointed out in Section\,4 a possible connection of the present subject 
to the study of quantum mechanical systems which have an underlying deterministic 
classical dynamical model \cite{tHooft01}. Indeed, we find that for certain 
``well-behaved time sequences'' the remaining deterministic aspects of the 
induced stochastic classical motion of the (in our model) external particle can be most 
simply described by a discrete-time quantum mechanical Schr\"odinger evolution. 
This ``stroboscopic quantization'' may arise under more general circumstances, if one 
insists to construct by statistical means an invariant ``time'' from localized observables.               

\subsection*{Acknowledgements} 
We thank M.\,Bahiana for computational advice and C.\,Wetterich for discussion. 
The work of H.-Th. Elze is partially supported under CNPq/DAAD 690164/01-7.  

\vspace{1.0cm}
\noindent\underline{\bf Figures:}

\begin{itemize}

\item[\underline{Fig.}\,1:] A typical trajectory. Initial 
conditions for the free motion on the torus are  
$\phi_{1,2}^{(0)}=0$ and $\pi_1^{(0)}=1$, $\pi_2^{(0)}=\sqrt 3$ (arbitrary units). 
Here $t=\tau$ (proper time); the final time is $\tau =50$.   

\item[\underline{Fig.}\,2:] The reparametrization invariant ``time'' $T$, Eq.\,(\ref{T}), 
based on localized incident counting, as a function of (computer) proper time $\tau$ 
for two different proper time intervals, cases a) and b); 
initial conditions as in \underline{Fig.}\,1; window parameter: $\delta =1/3$. 
The lines result from linear fits.  

\item[\underline{Fig.}\,3:] A typical trajectory with interaction parameters 
$\omega =\Omega =3$. The initial conditions are: 
$\phi_{1,2}^{(0)}=0$ and $\pi_1^{(0)}=3$, $\pi_2^{(0)}=\sqrt 5$; the final proper 
time ($t=\tau$) is $\tau =30$. 

\item[\underline{Fig.}\,4:] The reparametrization invariant ``time'' 
$T$, from Eq.\,(\ref{Ts}) with constant prefactor set to one, as a 
function of (computer) proper time 
$\tau$ for two different sets of initial conditions: $\phi_{1,2}^{(0)}=0$ (always), 
$\pi_1^{(0)}=5$, $\pi_2^{(0)}=\sqrt{17}$ (upper curve) 
and $\pi_1^{(0)}=3$, $\pi_2^{(0)}=\sqrt 5$ (lower curve); interaction parameters:  
$\omega =\Omega =3$, window parameter: $\delta =1/3$. Lines from linear fits. 

\item[\underline{Fig.}\,5:] Same as \underline{Fig.}\,4 (upper curve),  
illustrating nonlinear long-time behavior. 

\end{itemize}

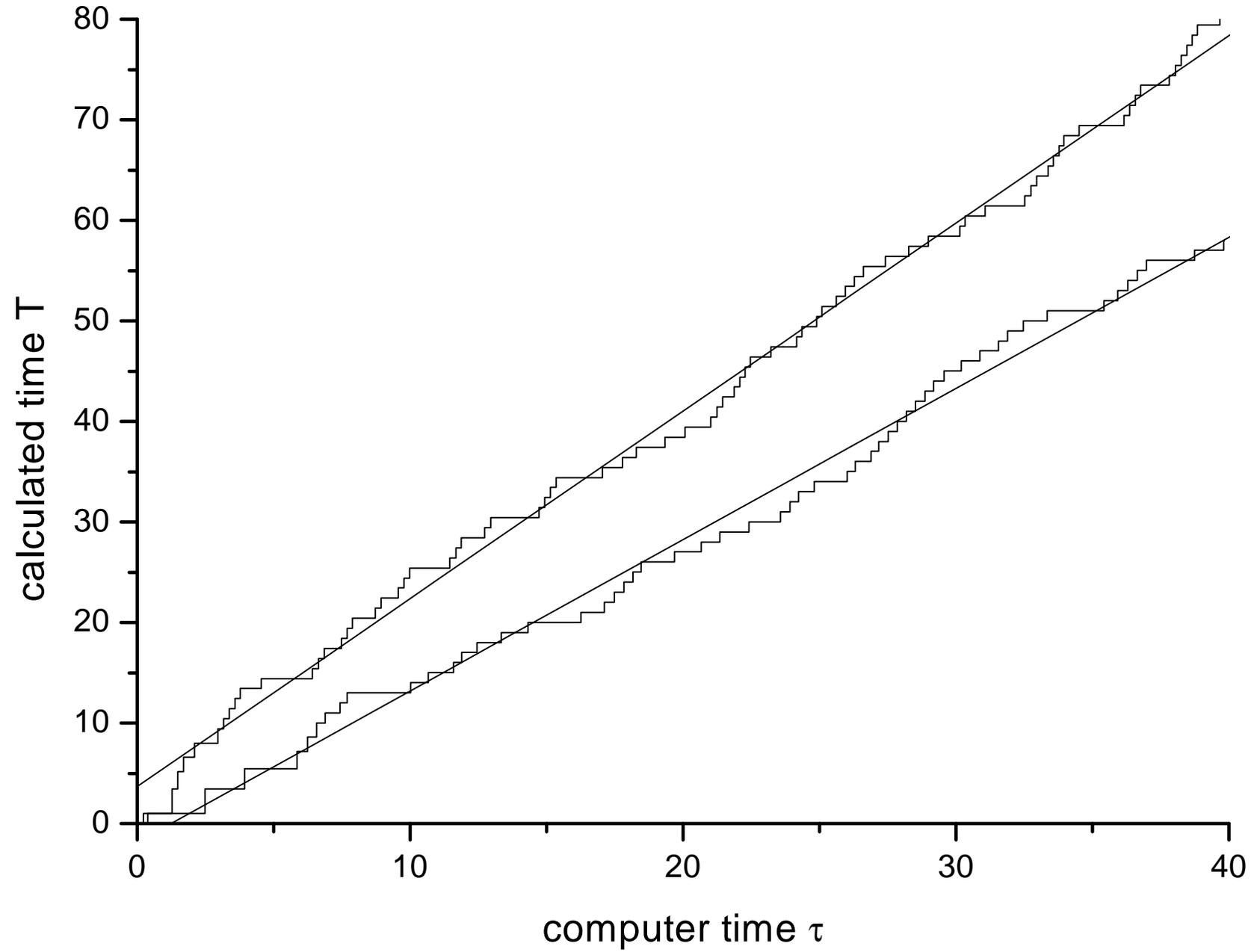
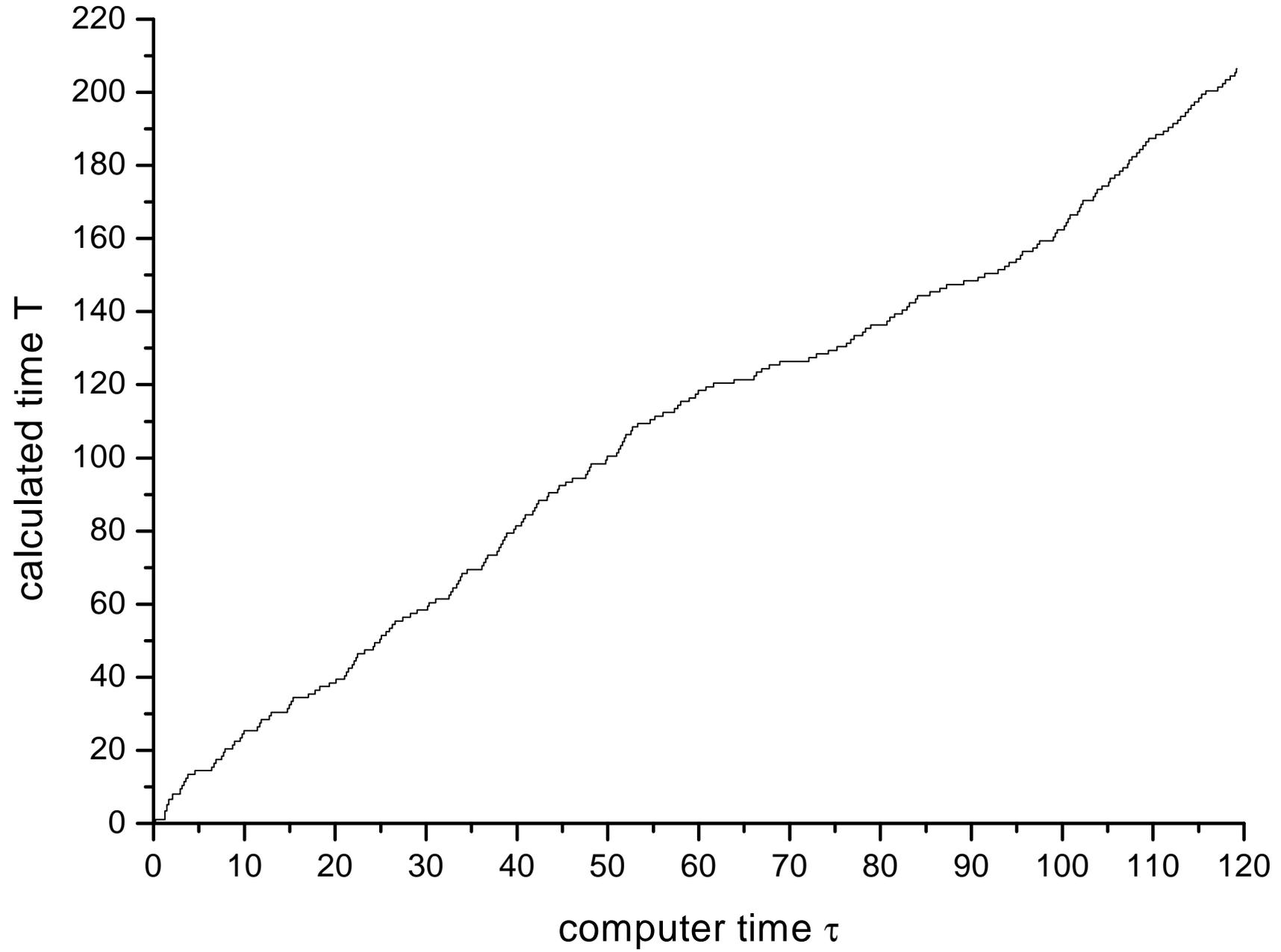
\end{document}